\documentstyle[aps,prl,floats,epsfig]{revtex}
\begin{document}
\twocolumn[\hsize\textwidth\columnwidth\hsize\csname@twocolumnfalse\endcsname
\title{Common Origin for Surface Reconstruction \\ 
	and the Formation of Chains of Metal Atoms}
\author{R.H.M. Smit, C. Untiedt, A.I. Yanson\thanks{Present address: Dept. of Physics, 510 Clark Hall, Cornell University, Ithaca, NY 14853} and J.M. van Ruitenbeek}
\address{Kamerlingh Onnes Laboratorium, Leiden University, Postbus
9504, 2300 RA Leiden, The Netherlands} \date{\today} 
\maketitle
\begin{abstract}
During the fracture of nanocontacts gold spontaneously forms freely suspended chains of atoms, which is not observed for the iso-electronic noble metals Ag and Cu. Au also differs from Ag and Cu in forming reconstructions at its low-index surfaces. Using mechanically controllable break junctions we show that all the 5$d$ metals that show similar reconstructions (Ir, Pt and Au) also form chains of atoms, while both properties are absent in the 4$d$ neighbor elements (Rh, Pd, Ag), indicating a common origin for these two phenomena. A competition between $s$ and $d$ bonding is proposed as an explanation. 
\end{abstract}
%\newpage
\vskip2pc]
\narrowtext
It has recently been discovered that nanowires of gold spontaneously evolve into chains of single atoms \cite{yanson98,ohnishi98}, which are surprisingly stable. They form metallic wires with a nearly ideal quantum value of the conductance, $G\simeq 2e^2/h$, and are able to sustain enormous current densities. Various numerical calculations on these chains have been presented, both in regular and distorted configurations \cite{chain_calculations}, but the question as to why these chains form specifically for Au, and, e.g., not for Cu or Ag, was not addressed \cite{preprint}. An understanding of the mechanism of the formation of these chains may help to improve our ability to control the fabrication process. This may lead to the formation of chains for different materials with interesting properties (magnetism, superconductivity) or of longer chains. Apart from a possible technical interest longer chains will also inspire new fundamental research as such atomic chains are the closest approximation to ideal one-dimensional metallic systems. These 1D systems are expected to undergo a Peierls distortion and ultimately Tomanaga-Luttinger Liquid effects
\cite{fisher97} could appear. 

In search for properties distinguishing Au from the other noble metals, that can be linked to the chain formation, we are particularly interested in surface effects, where the bonding between the atoms is modified by the reduced dimensions. Among the special features of Au that have been extensively studied are the reconstructions of the low-index surfaces. The  (110) surface shows a missing-row reconstruction, where every second row of atoms on the surface is removed; the (001) surface has a quasi-hexagonal reconstruction, where the top layer of the sample contracts to form a hexagonal layer on top of the square structure of the bulk. 

Since these surface reconstructions distinguish Au from Ag and Cu it is worth looking at the mechanism that has been put forward to explain them \cite{jacobsen}. In fact, the end-of-series 5$d$ elements Ir, Pt, and Au have similar surface reconstructions, which are absent in the related 4$d$ elements Rh, Pd and Ag, suggesting that the explanation for the reconstructions cannot lie in any particular detail of $d$ band electronic structure. There appears to be a growing consensus that a stronger bonding of low-coordination atoms of the 5{\it d} metals with respect to the 4{\it d} metals is a result of {\it sd} competition caused by relativistic effects in the electronic structure \cite{takeuchi89,pyykko88}. From the numerical work that uses relativistic local-density-functional calculations to evaluate the various contributions to the atomic binding energies qualitatively the following picture emerges. The effective Bohr radius $a_{0}=(4\pi \epsilon_{0})(\hbar^{2}/Zme^{2})$ for 1$s$ electrons of the heavier (5$^{th}$ row) elements contracts due to the relativistic mass increase $m=m_{0}/\sqrt{ 1-(v/c)^2}$. Higher $s$ shells also undergo a contraction, both because they have to be orthogonal against the lower ones and because they feel the same mass-velocity terms directly. As Takeuchi {\it et al.} explain \cite{takeuchi89} the contraction of the  $s$ shell reduces its energy, increasing the $s$ occupation at the expense of the $d$ electrons. 
Since at the top of the $d$ band the states with anti-bonding character are partially depleted, the $d$ bond becomes stronger. While the $d$ electrons thus tend to compress the lattice the $s$ electrons exert an opposing Fermi pressure. At the surface, the spill-out of the $s$ electron cloud into the vacuum relieves some of the electron pressure, and allows a contraction of the inter-atomic distance and a strengthening of the bonds at the surfaces, giving rise to the observed reconstructions. 

It is now natural to formulate the hypothesis that the consequences of these relativistic modifications in the metallic bonding will be even more pronounced for the one-dimensional chain structures. This suggests that we should look for possible chain formation in Pt and Ir, which have a reconstructed surface similar to Au, while we expect that Rh, Pd, and Ag, which do not reconstruct, should not show this tendency to chain formation. 

For reliable results on surface effects it is essential to prepare a clean sample surface. This can be achieved very simply with the use of Mechanically Controllable Break Junctions (MCBJ) \cite{ruitenbeek97}. Starting with a macroscopic wire, of purity better than 99.98\,\%, a notch is formed by incision with a surgical knife. For Rh and Ir due to the extreme hardness of the materials spark erosion was used to cut the notch. The samples are mounted inside a vacuum container and pumped overnight to a pressure below $5\cdot10^{-7}$ mbar. Then we cool to 4.2\,K in order to attain a cryogenic vacuum. Two hours after cooling the cryostat, once drifts due to thermal gradients are reduced to a minimum, the sample wire is broken at the notch by bending of the substrate onto which it has been fixed. The clean, freshly exposed fracture surfaces are then brought back into contact by slightly relaxing the bending. With the use of a piezo element the bending can then be fine-tuned to adjust the contact to atomic size.

A thick copper finger provides thermal contact to the sample inside the container. Helium thermal-exchange gas is avoided since adsorbed helium severely modifies the work function of the material \cite{kolesnychenko99} and the latter is used to calibrate the displacement of the two electrodes with respect to each other \cite{carlos}. Surface corrugation and various surface orientations slightly modify the effective tunnel barrier, giving rise to a 10\,\% uncertainty in the calibration. 

\begin{figure}[!t]
\begin{center}
    \leavevmode
    \epsfxsize=60mm
    \epsfbox{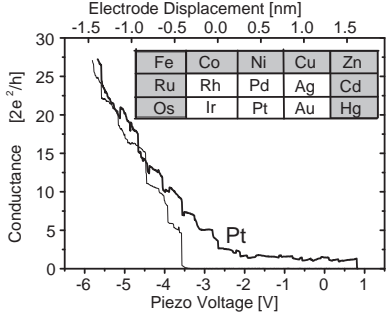}
\end{center}
\caption{\label{tracy}
Typical traces of the conductance as a function of the piezo voltage for Pt at 4.2\,K. The thick curve is recorded while breaking the contact, and the thin curve is taken while returning into contact. The formation of a chain of atoms can be recognized in the thick curve from the length of the last conductance plateau. The inset shows a fragment of the periodic table of the elements highlighting the transition-metals used in this study.}
\end{figure}

The contact is characterized by the conductance, measured with an accuracy of 1\%, at a bias voltage of 10--25\,mV. As the contact is slowly broken the conductance decreases in a step-wise fashion (Fig.~\ref{tracy}), where the last step value before the junction goes into tunneling typically corresponds to a contact of a single atom. We will use the length of the conductance plateau at the level of the conductance of a single atom as a measure for the length of the atomic chain being formed. 

\begin{figure}[!t]
\begin{center}
    \leavevmode
    \epsfxsize=60mm
    \epsfbox{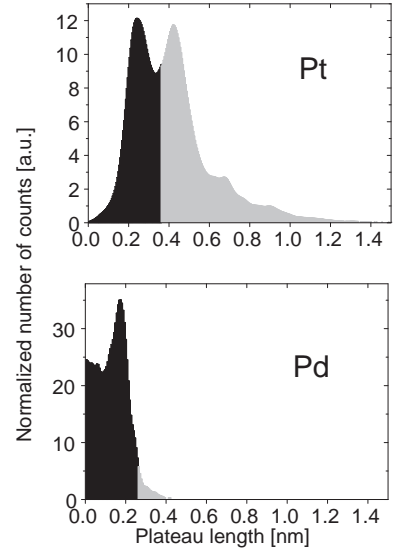}
\end{center}
\caption{\label{histo}
Histogram of the distribution of lengths for the last conductance plateau for platinum contacts (top panel) and for its iso-electronic neighbor palladium (bottom panel). Pd breaks essentially at the level of a single atom, while Pt spontaneously forms chains of atoms, which is apparent from the tail of the distribution towards long plateau lengths, and from the peaks at multiples of the inter-atomic distance. The area of the histogram above 1.5 times the chain length of the first peak is shaded gray, and defines the fraction $P_{1.5}$ used in Fig.\,\protect\ref{percentages}.}
\end{figure}

The formation of chains of metal atoms in the junction can be analyzed by the technique developed in Ref.~\cite{yanson98} for Au, which we briefly summarize. The distribution of the lengths of the last plateaus over a large number of contact-breaking cycles can be represented in a length-histogram (similar to the one for Pt in Fig.~\ref{histo}). As the conductance of a single atom of Au has been shown to be close to the fundamental unit of conductance $1\,G_{0}=2e^{2}/h$ \cite{scheer}, and since the conductance of the last plateaus is found very close to $2e^{2}/h$, the length of the plateau can be associated directly with the length of a chain of atoms. Strong support for this interpretation was found in the peaks of the length histogram. Their regular positions at multiples of 0.25~$\pm$~0.03~nm \cite{carlos} indicate that the average lengths at which the chain breaks comes in multiples of a length comparable to the atomic diameter (0.288\,nm for bulk Au), and agrees with the numerical calculations for the interatomic spacing in stable atomic chains
\cite{chain_calculations}. Further arguments are given in Ref.~\cite{yanson98}. Conclusive evidence for the formation of chains was obtained from the images of high-resolution TEM \cite{kizuka,rodrigues00,ugarte}, although the inter-atomic distances found here are usually bigger, between 0.3 and 0.4\,nm.

We now describe the application of the same MCBJ technique for Pt, followed by a description of the other transition metals in this study. Fig.~\ref{tracy} shows an example of a conductance trace for Pt. It shows a gradual decrease of the conductance, interrupted by sudden jumps when the atomic structure rearranges, as seen for most metals, followed by a long-stretched plateau around a reproducible conductance value equal to that expected for a single atom, and then a sudden break of the contact into the tunnel regime.
In contrast to Au, the last conductance plateau has a value well above 1\,$G_{0}$, and it is much less smooth and constant than the ones for Au. This is expected from the electronic structure for Pt, where the $sd$ orbitals provide up to 5 channels in the last atom \cite{scheer,cuevas}. The total conductance through these channels is strongly influenced by the local atomic configuration and the strain on the contact \cite{cuevas2}. 

\begin{figure}[!b]
\begin{center}
    \leavevmode
    \epsfxsize=50mm
    \epsfbox{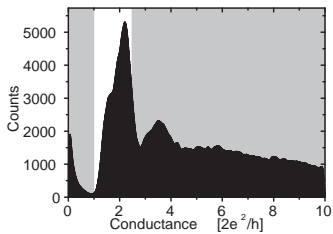}
\end{center}
\caption{\label{chisto}
Histogram of the number of times each conductance value is obtained for a large number ($\sim 2000$) of individual digitized conductance traces similar to the one shown in Fig.~\ref{tracy}, recorded while repeatedly breaking the Pt contact. The highlighted area indicates the range between the $G_{start}$ and $G_{stop}$ value used to define the length of the last plateau when recording the length-histogram in Fig.~\ref{histo}.}
\end{figure}

In order to determine experimentally the various possible conductance values for the last plateau we recorded a conductance histogram (Fig.~\ref{chisto}). The histogram is similar to the one in Ref.~\cite{thesis} and the dominant peak in this histogram is attributed to the conductance for a contact of a single atom \cite{ludoph00}. Based on this observation we start counting the length of a conductance plateau as soon as the conductance drops below a value, $G_{start}$, situated at the high end of the conductance peak in Fig.~\ref{chisto}, and we stop counting when it drops below a value, $G_{stop}$, at the low end of the peak. The two boundary values are illustrated in Fig.~\ref{chisto}. 
The histogram accumulated in this fashion for the plateau lengths in Pt is shown in Fig.~\ref{histo}(top). It shows a similar regular peak structure as was found
for Au. We emphasize that the shape of this histogram does not change significantly for different values of $G_{start}$ and $G_{stop}$. A similar histogram is found also for Ir \cite{iridium}. 
%The start and stop values used for the various metals are 
%summarized in Table~\ref{table I}. 
The plateau-length histograms give evidence that  Pt and Ir, similar to Au, form chains of atoms at the last stages of contact breaking.

When we apply the same method to Pd, however, we obtain a strikingly different histogram, with a much shorter tail towards long last plateau lengths than for Pt, Fig.~\ref{histo}(bottom). Here only one peak is clearly observable, at a much shorter length, and in contrast to Pt there is a high probability for breaking the contact at very short lengths. Similar results have been obtained for Rh and Ag. Clearly, Rh, Pd and Ag, which do not show the mentioned surface
reconstructions, also do not favor the formation of atomic chains.

\begin{figure}[!t]
\begin{center}
      \leavevmode
       \epsfxsize=60mm
        \epsfbox{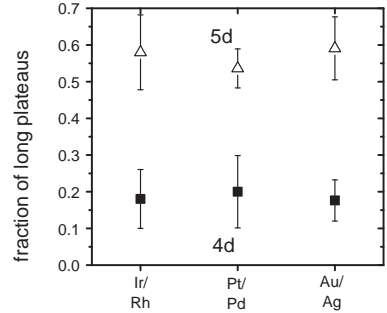}
\end{center}
\caption{\label{percentages}
The fraction $P_{1.5}$ of plateaus longer than 1.5 times the length at the first maximum in the length-histogram for the various metals, in the order of their position in the periodic table. The filled squares and open triangles indicate the fractions for the 4$d$ and 5$d$ metals, respectively. The error bars indicated in both graphs show the standard deviation of the result obtained by combining all histograms for the same material. These results show that only the 5$^{th}$ row elements have a preference to form long atomic chains.
The number of histograms recorded for each material, and the values $G_{start}$ and $G_{stop}$ used to determine the plateau lengths were 8, 3.4 and 1.0 for Rh; 6, 2.4 and 0.5 for Pd; 14, 1.1 and 0.5 for Ag; 6, 2.0 and 0.5 for Ir; 12, 2.5 and 1.0 for Pt; 12, 1.2 and 0.5 for Au. }
\end{figure}

In order to obtain an objective quantity that measures the tendency towards chain-configurations we define $P_{1.5}$ as the fraction of plateaus above 1.5 times the length of the first peak in the plateau-length histogram, as in Fig.\,\ref{histo}. The values obtained are shown in Fig~\ref{percentages} for all 4$d$ and 5$d$ metals considered in this study. The 5$d$ metals clearly have a significantly larger fraction of long last plateaus. The lengths obtained for the 4$d$ metals are dominantly in the range of lenths of only a single atom. The clear distinction between 4$d$ and 5$d$ metals in Fig~\ref{percentages} is not very sensitive to the precise definition for the fraction $P_{1.5}$. 

The results demonstrate a one-to-one correspondence between the metals showing the (110) missing-row and the (001) quasi-hexagonal reconstructions and those showing the formation of chains of atoms. This correlation only becomes meaningful when we can identify a common origin for the two properties. As for the reconstructions, the bonding in atomic chains will be influenced by a tilting of the balance between $s$ and $d$ electrons by relativistic effects. This difference in bonding becomes revealed when the Fermi pressure of the {\it s} electrons can be released by spill-out of the wave functions into the vacuum. As a result, there is a gain in energy from the stronger {\it d} bonds and a reduction of the inter-atomic distance. Clearly, the 1D chain geometry allows for an even larger {\it s} pressure release than at the surface. The effect of this can be recognized in the inter-atomic distances for the chains, obtained from the peak distances in the length histograms: 0.22$\pm$0.02\,nm for Ir, 0.23$\pm$0.02\,nm for Pt and 0.25$\pm$0.03\,nm for Au are all smaller than the bulk values (0.271, 0.277 and 0.288\,nm respectively). The important role of the {\it d} orbitals in the formations of Au atomic chains has recently been demonstrated by H{\"a}kkinen {\it et al.}\  using scalar-relativistic local density calculations \cite{chain_calculations}.

New structures in Au nanowires for larger diameters have recently been observed in HR-TEM experiments, in the form of helical multi-shell nanowires \cite{kondo00}. Numerical simulations for atomic-size metallic nanowires confirmed the stability of these structures for Au, but did not find them in Ag \cite{tosatti01}. This supports the
idea that properties of metal surfaces are strongly related to the behavior of
metallic nanowires. We propose that Pt and Ir are likely candidates to show similar weird wire helical structures \cite{gulseren98} as found for Au. From surface science we learn that surface reconstructions can be {\it induced} for Rh, Pd and Ag by deposition of a low density of specific adsorbates, which modifies the balance between the $s$ and $d$ bonds in the metal. It would therefore be interesting to test whether the 4$d$ metals can also be induced to form chains under the influence of similar adsorbates. Experiments to verify this are in progress.

This work is part of the research program of de ``Stichting FOM'', which is financially supported by NWO, and has been supported by a European Community Marie Curie Fellowship. We thank K.W. Jacobsen, S. Bahn, E. Tosatti and A.P. Sutton for informative discussions, and L.Y. Chen for assistance in the experiments.


\begin{references}
\bibitem{iridium} Iridium is very brittle and the highly corrugated electrodes that result after a break make it difficult to find a proper contact. It was necessary to frequently verify the calibration to obtain reliable results.
\end{references}
\end{document}